\documentclass[reprint, secnumarabic, amssymb, amsmath, aps, prl,cha, groupedaddress,frontmatterverbose,showpacs]{revtex4-1}
\usepackage[dvips]{graphicx}
\usepackage{bm}
\usepackage{color}

\begin{document} 
\title{Exchange striction induced  giant  ferroelectric polarization in copper based multiferroic material  $\alpha$-Cu$_2$V$_2$O$_7$.}
\author{J. Sannigrahi}
\author{S. Bhowal}
\author{S. Giri}
\author{S. Majumdar}
\author{I. Dasgupta}
\affiliation{Department of Solid State Physics, Indian Association for the Cultivation of Science, 2A \& B Raja S. C. Mullick Road, Jadavpur, Kolkata 700 032, INDIA}


\begin{abstract}
We report $\alpha$-Cu$_2$V$_2$O$_7$ to be an improper multiferroic with the simultaneous development of electric polarization and magnetization below $T_C$ = 35 K. The observed spontaneous polarization of magnitude 0.55 $\mu$Ccm$^{-2}$ is highest among the copper based improper multiferroic materials. Our study demonstrates sizable amount of magneto-electric coupling below $T_C$ even with a low magnetic field.  The theoretical calculations based on density functional theory (DFT) indicate magnetism in $\alpha$-Cu$_2$V$_2$O$_7$ is a consequence of {\em ferro-orbital} ordering driven by polar lattice distortion due to the unique pyramidal (CuO$_{5}$) environment of Cu. The spin orbit coupling (SOC) further stabilize orbital ordering and is crucial for magnetism. The calculations indicate that the origin of the giant ferroelectric polarization is primarily due to the symmetric {\em exchange-striction} mechanism and is corroborated by temperature dependent X-ray studies.  
\end{abstract}
\maketitle
\par
Recently multiferroic materials with mutually coupled ferroelectric (FE) and magnetic orders have attracted considerable interest for their versatile technological as well as fundamental importance.~\cite{eerenstein, cheong, fiebig, wang}  A strong magneto-electric (ME) coupling is expected in improper magnetic mutiferroics where ferroelectricity is induced by a specific magnetic order. In the last one decade, several magnetic multiferroics have been discovered ~\cite{kimura2,hur,lee, ni3v2o8,sg,sg1} where  
FE polarization is either associated with  {\it spiral magnetic structure} induced by spin-orbit coupling (SOC)~\cite{KNB, sergienko} or by {\it symmetric exchange striction} (SES) mechanism in case of collinear magnets.~\cite{cheongPRL, lee} Due to the secondary nature of the electric order,  the value of the FE polarization in such magnetic multiferroics is much  smaller (generally $\sim$ 0.01 $\mu$C.cm$^{-2}$) compared to the `proper' FE.~\cite{wang}
  A recent breakthrough in this direction is the discovery of giant ferroelectricity ($\sim$ 0.3 $\mu$C.cm$^{-2}$) and large ME coupling in mixed valent manganate CaMn$_7$O$_{12}$ below about 90 K~\cite{camn7o12exp} mediated by both Dzyaloshinski-Moriya (DM) interaction as well as exchange striction mechanism.~\cite{CaMn7O12} In this respect cuprates may be an attractive option as the orbital degrees of freedom and strong Coulomb correlations present in cuprates may not only lead to lattice distortion and magnetism but also possibly induce a coupling between them which are essential ingredients for multiferroicity.\\

\par
In view of the above, 
we investigated the Cu-based oxide Cu$_2$V$_2$O$_7$ in its orthorhombic $\alpha$ phase. Cu$_2$V$_2$O$_7$ crystallizes in at least three different polymorphs, namely $\alpha$, $\beta$ and $\gamma$-phases where only the $\alpha$ phase is non-centrosymmetric~\cite{calvo,alex,kriv} 
and is important in the present context. It consists of  magnetic Cu$^{2+}$ (3$d^9$, $S$ = $\frac{1}{2}$) and nonmagnetic V$^{5+}$ (3$d^0$, $S$ = 0) metal ions making it a system having both partially filled and empty $d$ shells similar to BiFeO$_3$, BiMnO$_3$, Pb(Fe$_{2/3}$W$_{1/3}$)O$_{3}$ etc.~\cite{catalan, kimura1}. All Cu$^{2+}$ ions are equivalent with fivefold coordination to oxygen atoms forming a distorted [CuO$_5$] polyhedron. Each Cu-polyhedron is linked with another two via edge sharing and they together form two sets of mutually perpendicular zig zag chains (see Fig. 1). These chains are separated by V$_2$O$_7^{4-}$ anionic group resulting from the two corner sharing VO$_4$ tetrahedra.~\cite{sanchez, calvo} 

\par
The magnetic behavior of $\alpha$-Cu$_2$V$_2$O$_7$ have been investigated earlier on polycrystalline samples.~\cite{sanchez, ponomarenko, Pommer}  It was reported that $\alpha$-Cu$_2$V$_2$O$_7$ is an antiferromagnet with weak ferromagnetism at low temperature. A magnetic order was seen below 35 K ($T_C$) accompanied by a change in slope in dielectric response near $T_C$. In this letter, a combined theoretical and experimental work establishes that the compound is an improper multiferroic with giant $P$ and ME effect where the origin of giant FE polarization is primarily due to symmetric exchange-striction mechanism, and the  magnetism is stabilized by  {\em ferro-orbital} ordering.

\begin{figure}[h]
\centering
\includegraphics[width = 8 cm]{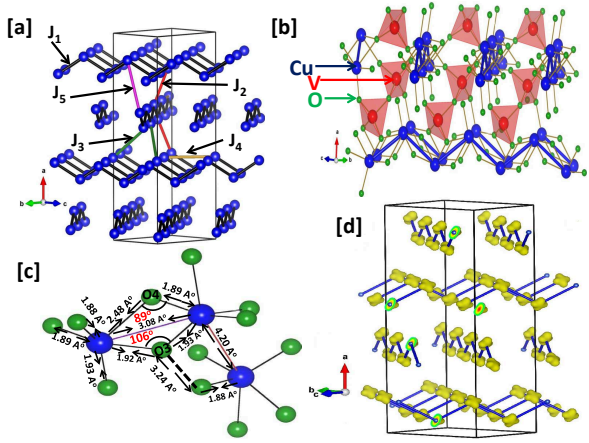}
\caption {(a) Cu atoms in the conventional unit cell form a pair of mutually perpendicular zig-zag chains. Various spin exchange interactions in $\alpha$-Cu$_2$V$_2$O$_7$ are also shown. (b) Edge sharing Cu-polyhedron forming two mutually perpendicular chains connected by two corner sharing VO$_4$ terahedra. (c) Change in the nearest neighbor Cu-O(4)-Cu and Cu-O(3)-Cu pathways as well as the  next nearest neighbor exchange path upon relaxation. (d) Three dimensional electron density plots showing the {\em ferro-orbital} order within LSDA+U+SOC.}
\end{figure}

\par
The experimental studies including magnetic, dielectric and electric polarization measurements were performed on a well characterized sintered polycrystalline sample,~\cite{reitveld} which have been described in detail in Supplemental Material (SM). All the electronic structure calculations presented in this paper are performed using DFT within local density approximation (LDA) and projector augmented wave (PAW) method as encoded in Vienna {\em ab-initio} simulation package (VASP)~\cite{PAW1,PAW2,vasp1,vasp2} (see SM). 
\par
Fig. 2 (a) describes  magnetization ($M$) {\it vs.}  temperature ($T$) data  in  zero-field-cooled-heating (ZFCH), field-cooling (FC) and field-cooled-heating (FCH) protocols under magnetic field $H$ =  100 Oe.  $M (T)$ shows a sharp rise at $T_C$ = 35 K indicating  the transition to a magnetically ordered state. The thermal hysteresis between FC and FCH around $T_C$ indicates the first order nature of this transition. The inverse molar susceptibility ($\chi^{-1}(T)$, where $\chi = M/H$) obeys  Curie-Weiss law above 80 K (see inset of  fig.2(a)) and we get Curie-Weiss temperature ($\theta_C$) to be $\approx$ $-$78 K which indicates the predominant antiferromagnetic (AFM) correlations in the system. The effective moment of Cu$^{2+}$  is $\approx$ 1.92 $\mu_B$ and it is slightly higher than the spin-only value (= 1.73 $\mu_B$).  At low $T$, $M$ almost saturates which is not a likely  behavior of a pure AFM ordering. Possibly, the magnetic state below $T_C$ is canted AFM type. 
The isothermal $M$ {\it vs.} $H$ at 5 K for $H$ = $\pm$ 9 kOe is shown in the main panel of Fig. 2 (b). The curve shows clear hysteresis which reaffirms the presence of ferromagnetic (FM) component.  The coercivity of the loop is found to be about 2 kOe. The full loops both at 5 K and at 150 K (well above $T_C$) are shown in the inset. The $M-H$ curve at 5 K, however, does not show full saturation even at 50 kOe, and it once again indicates canted spin structure. We can fit the high field data (between $H$ = 30-50 kOe) with an empirical relation $M (H) =  \chi_{afm}H + M_S$, where $\chi_{afm}H$ is the linear term due to AFM component and $M_S$ is the saturation magnetization due to the FM part. We find $M_S$ to be 0.08 $\mu_B$/f.u., which is quite small compared to the full saturation moment of two Cu$^{2+}$ ($\sim$ 2 $\mu_B$) ions indicating the presence of weak ferromagnetism. Interestingly, the 5 K isotherm is not found to be quite smooth, and it contains signature of sharp jump whenever the field changes its sign. This may indicate the presence of uniaxial anisotropy in the system.

\par
Fig. 3 (a) shows the $T$-variation of the real part of the complex dielectric permittivity ($\epsilon^{\prime}$) measured at different frequencies ($f$). $\epsilon^{\prime}$ is almost constant and independent of $f$ in the low-$T$ regime (below $\sim$ 130 K) which signifies the static dielectric constant originating from the intrinsic contribution~\cite{ccto}. A closer look at low-$T$ part shows the existence of a small but clear hump-like anomaly around 35 K coinciding with $T_N$.
It is free from any frequency dispersion, suggesting that this feature may be related to some long range electric order.  
The imaginary part of  permittivity is quite small (particularly below about 130 K) indicating that the sample is highly resistive and the estimated resistivity at 50 K is found to be $\sim$ 100 M$\Omega$-cm.

\begin{figure}[t]
\centering
\includegraphics[width = 8 cm]{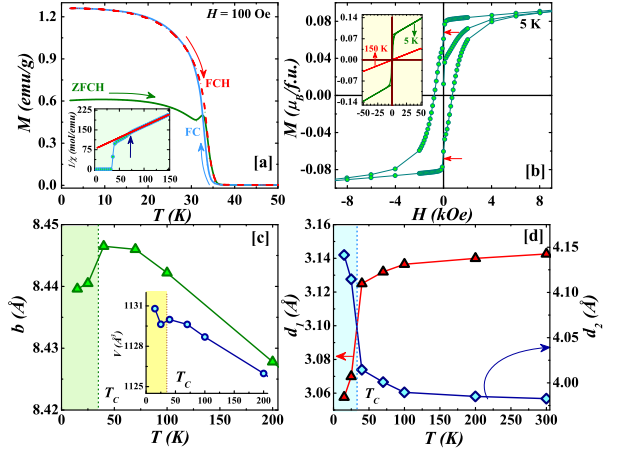}
\caption {(a) The $T$ dependence of ZFCH, FC and FCH magnetization data  of $\alpha$-Cu$_2$V$_2$O$_7$. The  inset shows  1/$\chi$ versus $T$ along with the fitting of Curie-Weiss law. (b) indicates the isothermal $M$ vs $H$ data  at 5 K and 150 K. The main panel of (c) shows the $T$ variation of orthorhombic lattice parameter $b$ along with inset depicting the $T$ dependence of lattice volume $V$ obtained from powder XRD data. (d) shows the change in bond lengths $d_1$ and $d_2$ corresponding to the interactions $J_1$ and $J_2$ (see fig. 1 (a)) with $T$. The inset shows change in the CuO$_{5}$ polyhedron after relaxation as obtained from theory.}
\end{figure}

\par
Considering the electric anomaly near the magnetic transition, it is tempting to measure the magneto-dielectric properties of the sample~\cite{bimno3}. Fig. 3 (b) shows the $T$ variation of $\epsilon^{\prime}$ measured  at $H$ = 0 and 9 kOe. Clearly, $\epsilon^{\prime}$ shows significant effect of magnetic field below about 80 K. In the inset of fig. 3 (b), we have plotted the change in $\epsilon^{\prime}$ as a function of $T$ due to the application  9 kOe of field and we observe a significant value of magneto-dielectric effect (as high as 3.5\%) at around 30 K. This is remarkably large considering the small value of the applied field .

\par
In order to shed more light on the nature of the hump-like feature observed in the $\epsilon^{\prime}$($T$) coinciding with the magnetic anomaly at $T_C$, we measured pyroelectric current ($I_P$) (see SM) after cooling the sample from room temperature with different electric field ($E_{Cool}$). From the $T$ variation of $I_P$, spontaneous polarization has been calculated (see fig 3 (c)), which we denote by $P_I$. Clearly $P_I$ shows a sharp increment below  about 35-40 K, eventually saturating at a lower $T$. 
The magnitude of $P_I$ clearly increases with the cooling field. We also measured $P_I$ with different polarity of $E_{Cool}$ and $P_I$ changes sign depending on the chosen sign of $E_{Cool}$ (see inset of fig. 3 (c)). Such behaviors of $P_I$ confirm that the sample undergoes long range FE order below 35 K with the development of spontaneous polarization. Since, the electric order is concomitant with the magnetic order, the sample can be assigned as an improper multiferroic material. It is to be noted that even at room temperature  $\alpha$-Cu$_2$V$_2$O$_7$ possesses a non-centrosymmetric crystal structure with polar point group ($mm2$), which in general belongs to the pyroelectric class of crystals.~\cite{noncentro} However, a switchable spontaneous $P_I$ is only achieved  below $T_C$, possibly arising from the favorable lattice distortion associated with the magnetic order. Remarkably, from the pyroelectric measurement the saturation value of the spontaneous $P_I$ is found to be  as large as  0.55 $\mu$C.cm$^{-2}$, which is substantially high compared to the other copper based magnetically driven ferroelectrics to date.~\cite{ishiwata, kimuranature} 

\par
The ferroelectricity is further confirmed by the measurement of  electric polarization versus electric field ($P-E$) loop using a FE loop tracer as shown in Fig. 3 (d).  The data recorded above $T_C$  (50 K and 80 K) do not show any loop, whilst clear loops with tendency for saturation are observed at 10 K and 30 K (which are  below $T_C$). Observation of such prototypical hysteresis loop in $P$ is an essential proof for the development of FE state below $T_C$.  At the value of $E$ = 1 kV.cm$^{-1}$ hysteresis almost closes and we get a value of polarization close to  1 $\mu$C.cm$^{-2}$.

\par
The experimental results discussed above lend support to the fact that ferroelectricity is induced by magnetism in $\alpha$-Cu$_2$V$_2$O$_7$ resulting in substantially large magneto-dielectric response. It is also interesting to note that the magnetic transition at $T_C$ is first order in nature which indicates possible structural transition associated with the magnetic as well as electric orderings. In view of the above, both  exchange striction as well as inverse DM effect may be the likely mechanism for the giant FE polarization in $\alpha$-Cu$_2$V$_2$O$_7$.  The  DM interaction is not unlikely here as the obtained effective moment per Cu site (1.92 $\mu_B$) is bit higher than the spin only moment (1.73 $\mu_B$) presumably due to the  orbital contribution of  the magnetic moment as a consequence of finite SOC.  It is to be noted that the spontaneous polarization in $\alpha$-Cu$_2$V$_2$O$_7$ is about one order of magnitude higher than the other spiral DM type multiferroics (such as TbMnO$_3$, where $P \sim$ 0.05$\mu$Ccm$^{-2}$). Therefore, the exact origin of FE state in this compound may involve more complex mechanism.

\par
To understand the electronic structure, the exchange mechanism and the origin of ferroelctric polarization, we have performed first principles DFT calculations, using VASP as described in SM. The non-spin polarized electronic structure calculations reveal that the oxygen $p$-states are completely occupied while the vanadium-$d$ states are empty and the Fermi level is hosted  by half-filled predominantly Cu-$d_{x^2-y^2}$ states consistent with the Cu$_2^{2+}$V$_2^{5+}$O$^{2-}_7$ nominal ionic formula for the system.  The distortion of the CuO$_5$ polyhedra is triggered by the orbitally active Cu$^{2+}$ ion in such a way that Cu-$d_{x^2-y^2}$ states are well separated from the rest of the Cu-$d$ states thereby promoting ferro-orbital ordering.  Such a ferro-orbital ordering will favor anti-ferromagnetic coupling between the Cu ions in the chain.~\cite{khomskii} It is also interesting to note that the distortion is such that the nearest neighbor (NN) Cu ions in each chain are now coupled by two asymmetric bonds Cu-O(3)-Cu and Cu-O(4)-Cu forming a non-centrosymmetric CuO$_2$ plaquette (see Fig. 1(c))  which will not only favor DM interaction but also stabilize  ferroelectric polarization. \\

To account for the observed magnetism in this system, we have evaluated various symmetric spin exchange interactions between the Cu-ions by performing total energy calculations in the framework of LDA + $U$ method.~(see SM) The value of $U_{eff}$ ($U-J$) was taken to be 6.5 eV following the usual choice for the cuprates.~\cite{U} Constraining the range of interaction to 5.42 \AA, we calculated five dominant exchange interactions~\cite{exchange} (the various spin-exchange paths are  shown in Fig. 1(a)). The dominant inter-chain exchange interaction  $J_3$ (-13.61 meV) is antiferromagnetic, followed by intra-chain  $J_1$(-4.67 meV)  and interchain $J_2$ (4.07 meV)  which are AFM and FM respectively.  Other exchange interactions $J_4$(0.26 meV)  and $J_5$ (2.37 meV) are small and FM. The AFM exchange interactions $J_1$  and $J_3$ are mediated via Cu-O-Cu and Cu-O-V-O-Cu paths respectively [See Fig. 1(b)]. The larger bond angles in the $J_3$ exchange path make this interaction stronger compared to $J_1$. The NN AFM $J_1$ is consistent with the {\em ferro-orbital} order and the strong inter-chain coupling  $J_3$ identified by our first principles calculations is responsible for the long range magnetic order. The FM exchange interaction $J_2$ is mediated  by the exchange path Cu-O-O-Cu where the two O atoms that mediate the exchange interaction between Cu ions are at an angle ($\angle${Cu-O-O}) of 
70.07$^{\circ}$ thereby favoring FM interaction. Using the computed exchange parameters, the Curie-Wiess temperature $\theta_{C}$ for $\alpha$-Cu$_2$V$_2$O$_7$ is calculated to be -77.4 K which is remarkably close to the experimental value (-78 K).\\

\begin{figure}[t]
\centering
\includegraphics[width = 8 cm]{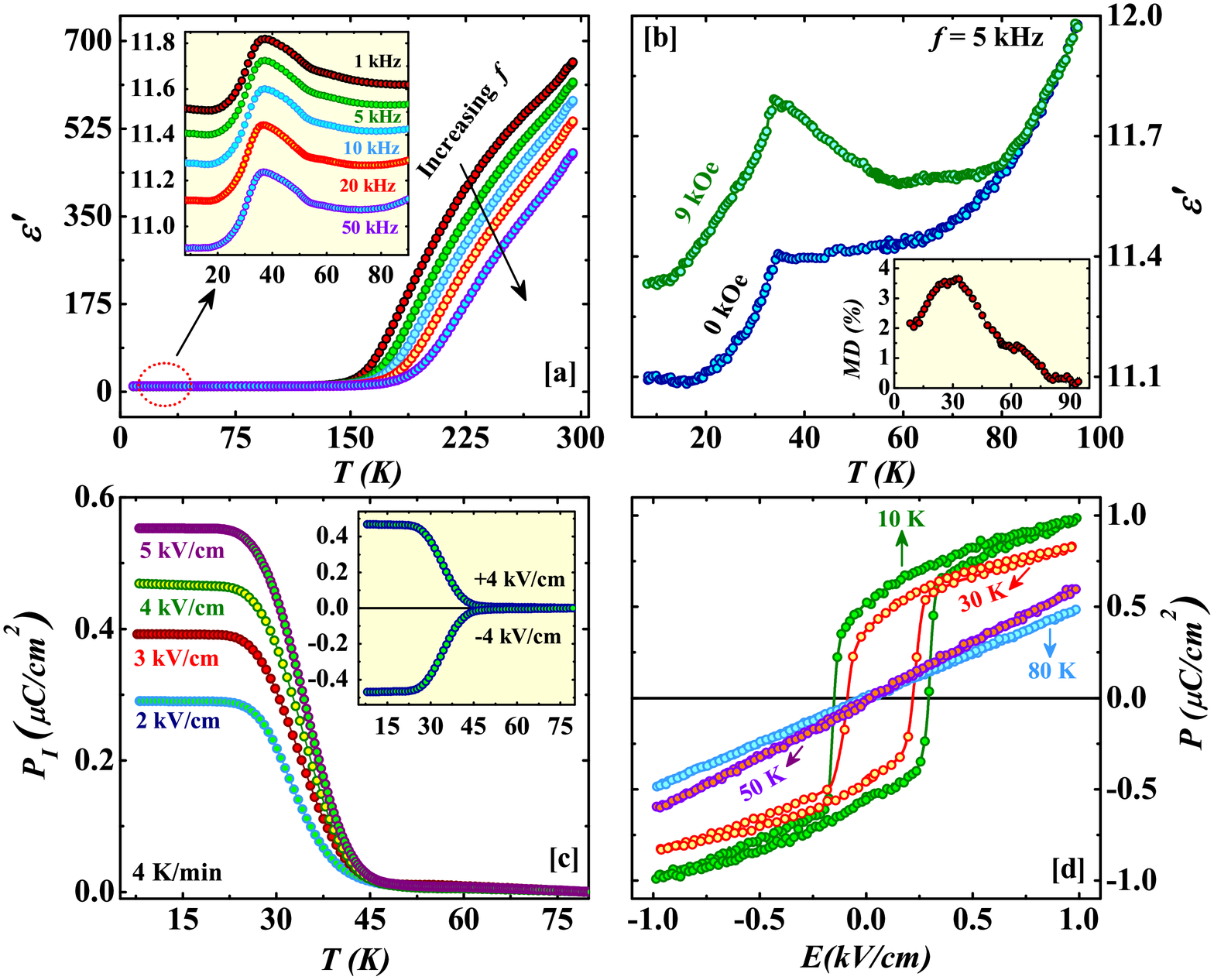}
\caption {(a) shows the $T$ variation of  the real part of dielectric permittivity.(b) shows the  $\epsilon^{\prime}$ data measured at $H$ = 0 and 9 kOe  along with the inset showing the change in $\epsilon^{\prime}$ due to $H$. (c) represents the  temperature dependence of   spontaneous electric polarization, $P_I$ calculated from the pyroelectric current measurements. Insets of the (c)  represents $P$ with positive and negative cooling electric fields. (d) shows the polarization hysteresis ($P-E$) loop measured at different temperatures.}
\end{figure}

\par
In the absence of any spin frustration, the canted AFM suggested by the $M-H$ curve may be attributed to the DM-type interaction due to SOC. We have therefore considered the antisymmetric part of the spin Hamiltonian $\mathcal{H} =\sum_{ij} \vec{D_{ij}} \cdot (\vec{S_i} \times \vec{S_j})$ and calculated the three components  $D^x$, $D^y$, $D^z$ of the DM parameter up to 3rd nearest neighbor for $\alpha$-Cu$_2$V$_2$O$_7$  by performing LDA + $U$ + SOC calculations. The ratio $|\frac{\vec{D}_{i}}{J_i}|$ $\sim$0.5 suggests unusually strong DM interaction in $\alpha$-Cu$_2$V$_2$O$_7$ very similar to that reported for CaMn$_7$O$_{12}$ system \cite{CaMn7O12}. The calculation of magnetocrystalline anisotropy energies reveal $b$-axis of the conventional unit cell to be the easy axis. \\
\par
Guided by the exchange interactions and in view of the importance of the SOC, we have considered three magnetic configurations namely  FM, AFM and non-collinear (NONC) as shown in Fig. 6 of SM.  In our model NONC magnetic configuration, the nearest neighbor spins in each zig-zag chain are antiparallel in the b-direction with a very small component along a and c direction as a result of canting. Our calculations (see Table-II of SM) reveal SOC stablizes magnetism  and the non-collinear magnetic structure is  energetically degenerate  with a very small magnetic moment(0.17$\mu_B$/f.u.). A three dimensional plot of the spin density for the NONC structure with LDA+U+SOC  confirm {\em ferro orbital} ordering (see Fig.1 (d)) where the same orbital is occupied at each site but rotated with respect to each other due to the distorted CuO$_{5}$ polyhedron.~\cite{OO1, OO2} \\

\begin{table} [h]
\caption{Calculated Polarization in various magnetic structures}
\centering
\begin{tabular}{c c c c }
\hline\hline 
  Config. \ & $\Delta E$ \ &  Polarization due to \ &  Polarization due to \\[0.3 ex]
           \ &   (meV)    \ & SOC                  \ & Exchange striction    \\
            \ &            \ & $\Delta P_{SOC}$     \ & $\Delta P_{ex}$       \\
             \ &            \ &  ($\mu$C.cm$^{-2}$)  \ & ($\mu$C.cm$^{-2}$)   \\
\hline
FM  &  29 & 0.02 & 3.72 \\
AFM &  0 & 0.05 & 4.08  \\
NONC & 0 & 0.01 & 4.03  \\
\hline
\end{tabular}
\end{table}
\par
Finally the FE polarization is calculated for the above mentioned three magnetic configurations using the Berry phase method \cite{berry} as implemented in VASP.\cite{vasp1} For the purpose of  understanding the contribution of SOC and {\em exchange-striction} on ferroelectric polarization, calculations are carried out including SOC both for the experimental structure and the  relaxed structure. The results of our calculation (Table I) although suggest the importance of SOC but conclusively establish that {\em exchange-striction} is the primary mechanism for the giant ferroelectric polarization for this system. As shown in Fig. 2(d) (inset), as a result of exchange striction there is a compression of the CuO$_{5}$ polyhedron, resulting in shorter bond length in the exchange path $J_{1}$ and consequently increase in the bond length in the exchange path $J_{2}$ (See Table II of SM). Shorter bond length $d_1$ in the AFM exchange path $J_1$ will enhance the hopping contribution $t_1$ and add to the gain in energy by  the superexchange ($\propto$ $t_1^{2}/U$).
\par
 In order to corroborate our theoretical results on {\em exchange-striction} we performed $T$-dependent X-ray diffraction (XRD) measurement on the sample . Our analysis of the data indicates that the crystal symmetry ($Fdd2$) remains unchanged both below and above $T_C$.~\cite{neutron}  However, there is clear  change in the orthorhombic lattice parameters ($a$, $b$ and $c$)  at $T_C$, where the change is prominent for lattice parameter b (fig. 2 (c)). Remarkably  on cooling across $T_N$ there is a sharp first order like change in $d_1$ and $d_2$ (see fig. 2 (d)) with $\Delta d_1$ = -0.085(2) \AA  and  $\Delta d_2$ = 0.159(2) \AA, while $d_3$ almost remains unchanged ($\Delta d_3$ =-0.01 \AA ) in excellent agreement with our theoretical prediction. (See Table II of SM)      

\par
In conclusion, we have found that $\alpha$-Cu$_2$V$_2$O$_7$ is a magnetic multiferroic material with the highest FE polarization among the known Cu based multiferroic oxides with sizable amount of ME coupling. $\alpha$-Cu$_2$V$_2$O$_7$ turns out to be an unique example of multiferroic material with single valent Cu$^{2+}$ ions  where orbital degrees of freedom lead to polar distortion and {\em ferro-orbital} ordering  favoring  AFM. The SOC further assists to stabilize orbital ordering and magnetism. Finally the AFM interaction promotes large {\em exchange-striction} and is the primary mechanism that gives rise to giant electric polarization.
\par
The work is supported by the grants from BRNS, India (2012/37P/39/BRNS/1991). We thank low temperature XRD facility, ECMP division, SINP, Kolkata for $T$ dependent XRD measurements.

{99}

\newpage

\vspace{.5 cm}
\begin{center}
\bf{Supplementary Materials (SM) for\\
\vspace{.3 cm}
Exchange striction induced  giant  ferroelectric polarization in copper based multiferroic material  $\alpha$-Cu$_2$V$_2$O$_7$}
\end{center}

\section{Experimental Aspects}
\subsection{Sample Preparation and Characterization}
Polycrystalline sample of Cu$_2$V$_2$O$_7$ was prepared by conventional solid state reaction route in air. Highly pure CuO and V$_2$O$_5$ were mixed thoroughly in a stoichiometric ratio and homogenized with ethanol in an agate mortar. The mixture was pressed into pellets and sintered at 600$^{\circ}$ C for 90 h with several intermediate grindings. We found that any high temperature sintering results in the formation of the $\beta$ phase of Cu$_2$V$_2$O$_7$. Powder X-ray diffraction (XRD) patterns at room temperature (300 K) as well as at 200 K, 100 K, 70 K, 40 K, 25 K and 15 K  were recorded  using Cu K$\alpha$ radiation in Rigaku-TTRAX-III diffractometer which ensure the single phase of the polycrystalline sample with no detectable secondary phase.  Reitveld refinement analysis were performed  with the help of MAUD software on the powder XRD patterns. The XRD patterns at all temperatures (both above and below $T_C$ = 35 K)  are well fitted in orthorhombic crystal structure with Fdd2 space group where reliability factors ($\sigma$) are $<$ 1.4. The lattice parameters at 300 K were found to be $a$ = 20.692 \AA, $b$ = 8.413 \AA and $c$ = 6.451 \AA. The Cu-O-Cu and V-O-V angles are  found to be 106$^{\circ}$ and 148$^{\circ}$ respectively. These angles as well as Cu-O and V-O bond lengths agree well with the previous work on the structure of $\alpha$-Cu$_2$V$_2$O$_7$.~\cite{calvo1} Although no change in crystal symmetry was observed (within the resolution of our data), clear change in lattice parameters and bond lengths are present around the Curie point.
The deviations in the lattice parameters and in the bond lengths with temperature are well outside the error limit of refinement which are respectively $\approx$ 0.0006 \AA and $\approx$ 0.002 \AA.

\subsection{Magnetic measurements}
The magnetic measurements were performed on a Quantum Design SQUID magnetometer (MPMS-4, Evercool). Extra precautions were taken to judge the small thermal hysteresis observed in the isothermal magnetization measured at 5 K where data were recorded with stable mode option.

\subsection{AC dielectic measurements}
The ac dielectric measurements were performed using an Agilent E4980A precision LCR meter in the temperature range 5-300 K in a helium closed cycle refrigerator. An electromagnet with maximum field strength of 9 kOe (at 5 cm pole separation) was used to apply magnetic field during dielectric measurements.

\subsection{Pyroelectric Measurement}
The pyroelectric current of the sample was measured using a Keithley Electrometer (model 6517B) in the helium closed cycle refrigerator. From the pyroelectric current polarization have been  calculated (see fig. 4). In order to record $I_P$, a capacitor type arrangement was used where a pair of electrodes were attached to two flat surfaces of the pelletized sample using silver epoxy. The sample was first cooled down to 10 K in presence of electric field ($E_{Cool}$ = 2, 3, 4 and 5 kV/cm). After reaching lowest $T$,  $E_{Cool}$ was set to zero and $I_P$ was measured during heating of the sample at a constant rate of 4 K/min.~\cite{inomata, kitamura}  The pyroelectric current density($J_P$) was calculated by dividing $I_P$ by the area of the electrode($A$). We can calculate $P$ with the following relation: $$P_I = -\frac{1}{A{\left(\frac{dT}{dt} \right )}}\int_{T_1}^{T_2}I_PdT$$, where $\frac{dT}{dt}$ is the rate of change of temperature. Here we assume that $P_I$ vanishes above 45 K.  Clearly a peak is noticed around 35 K and the value of $J_P$ increases with the increase of $E_{Cool}$. The peak position is almost cooling field independent which indicates that the intrinsic pyroelectric current is dominant.

\begin{figure}[ht]
\centering
\includegraphics[width = 8 cm]{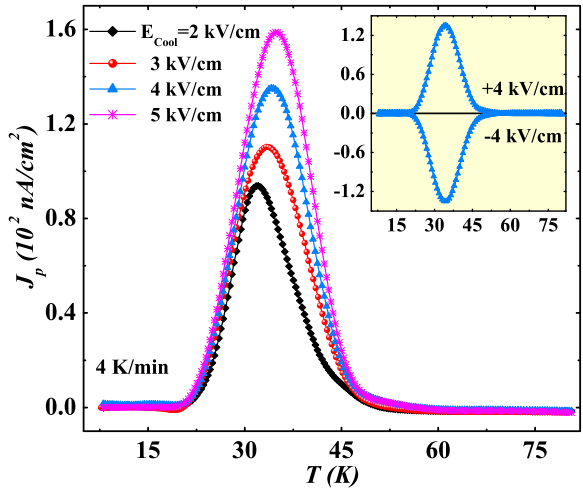}
\caption {$T$ dependent  pyroelectric current density of $alpha$-Cu$_2$V$_2$O$_7$ measured after cooling the sample in different electric fields.}
\end{figure}

\begin{figure}[ht]
\centering
\includegraphics[width =8 cm]{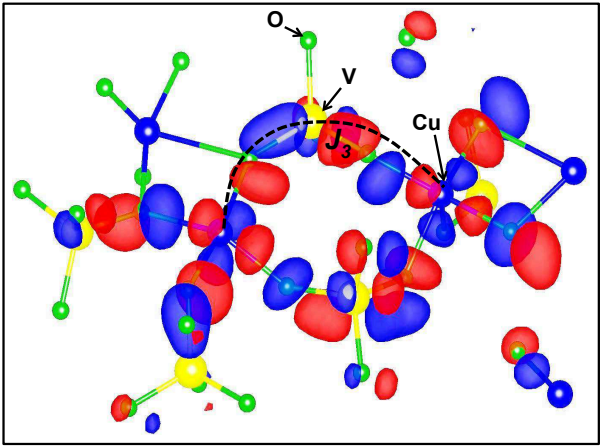}
\caption {Effective  Cu-d$_{x^2-y^2}$ Wannier function plot showing the exchange path for the inter-chain exchange interaction J$_3$.}
\end{figure}

\begin{figure}[ht]
\centering
\includegraphics[width =8 cm]{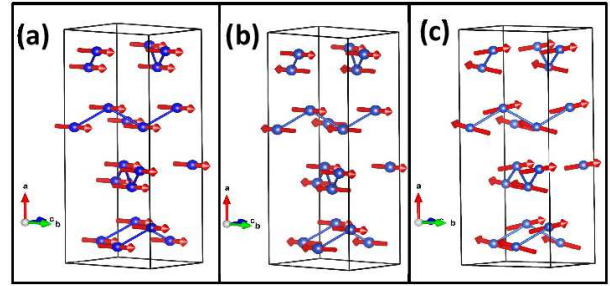}
\caption {Different magnetic ordering (a) ferromagnetic (FM), (b) antiferromagnetic (AFM), (c) noncollinear (NONC).}
\end{figure}

\subsection{Polarization hysteresis loop measurement}
Polarization hysteresis loops ($P-E$) at different constant temperatures were measured by a Ferroelectric Loop Tracer from Radiant Technology (Precision Premier-II) in a helium closed cycle refrigerator. Measurements were performed in presence of electric field as high as $\pm$ 1 kV/cm with a time period of 10 ms in standard bipolar mode.

\section{Theoretical Aspects}
\subsection{General techniques}
All the electronic structure calculations presented in this paper are performed using first principles density functional theory (DFT) within local density approximation (LDA) including Hubbard U \cite{hubu} using projector augmented-wave (PAW) method \cite{pawm1,pawm2} encoded in the Vienna {\it ab initio} simulation package (VASP) \cite{vasps1,vasps2}. The values of on-site Coulomb interaction (U) and Hund’s rule coupling ($J_H$) parameters were taken as $U$=7.5 eV, $J_H$=1.0 eV \cite{U}. The energy cutoff for the plane wave expansion of the PAW’s was taken to be 550 eV. A (4$\times$4$\times$4) k-mesh has been used for self consistency. Symmetry has been switched off in order to minimize possible numerical errors.
The calculation for the ferroelectric polarization with FM, AFM and NONC magnetic configuration are carried out using Berry phase method \cite{berryphase} as implemented in the Vienna {\it ab initio} simulation package (VASP). The Wannier function for the Cu-d$_{x^2-y^2}$ orbital (shown in Fig. 5) is constructed using the
VASP2WANNIER and the WANNIER90 codes.~\cite{w90}  \\

\vspace{.25 cm}
\begin{table} [ht]
\caption{The relative energies and the change in Cu-Cu bond lengths upon relaxation for the nonmagnetic (NM) and different magnetic configurations with and without SOC have been listed here. + (-) signs indicate the increment (decrement) of the bond-length.}
\centering
\begin{tabular}{|c |c |c |c |c |c | c|}
\hline
Structure  & $\Delta E$ & \multicolumn{5}{|c|}{Change in bond lengths  }   \\[0.2 ex]
          \ & (meV) & \multicolumn{5}{|c|}{upon relaxation  corresponding} \\
          \ &  \ & \multicolumn{5}{|c|}{ to the following } \\
          \ &  \ & \multicolumn{5}{|c|}{exchange paths with respect to }\\
          \ &   \ & \multicolumn{5}{|c|}{the experimental structure (\AA)} \\
          \ &    \ & J1 \ &  J2 \ &  J3 \ &  J4 \ &  J5   \\
\hline
NM+U relax & 130 & 0.0 &   -0.01&  0.0 &  0.0& 0.0 \\
FM+U relax & 120 &-0.07& 0.25& -0.01& 0.0& 0.0 \\
AFM+U relax & 82 & -0.06& 0.22 & -0.01& 0.0& 0.0 \\
NONC+U relax & 80 & -0.07 & 0.23 & -0.01& 0.0& 0.0\\
FM+SOC+U relax & 29  & -0.07& 0.25& -0.01& 0.0& 0.0 \\
AFM+SOC+U relax & 0 &-0.07& 0.24 & -0.01& 0.0& 0.0 \\
NONC+SOC+U relax & 0 &-0.07 & 0.23 & -0.01& 0.0& 0.0\\[1ex]
\hline
\end{tabular}
\end{table}



\begin{thebibliography}{99}
\bibitem{exchange} The various exchange interactions are designated depending on the distance between the Cu atoms. Here J$_{1}$ and J$_{4}$ are intra-chain interaction while J$_{2}$, J$_{3}$, J$_{5}$ are inter-chain exchange interaction.
\end{thebibliography}
\end{document}